\documentclass[a4paper,english,aps,preprint]{revtex4}
\usepackage[T1]{fontenc}
\usepackage[latin9]{inputenc}
\usepackage{color}
\usepackage{graphicx}
\usepackage{esint}

\makeatletter

\providecommand{\tabularnewline}{\\}

\@ifundefined{textcolor}{}
{%
 \definecolor{BLACK}{gray}{0}
 \definecolor{WHITE}{gray}{1}
 \definecolor{RED}{rgb}{1,0,0}
 \definecolor{GREEN}{rgb}{0,1,0}
 \definecolor{BLUE}{rgb}{0,0,1}
 \definecolor{CYAN}{cmyk}{1,0,0,0}
 \definecolor{MAGENTA}{cmyk}{0,1,0,0}
 \definecolor{YELLOW}{cmyk}{0,0,1,0}
 }

\makeatother

\makeatother

\usepackage{babel}

\begin{document}

\preprint{}

\title{Coupled thermo-mechanics of single-wall carbon nanotubes}

\author{F Scarpa}

\email{f.scarpa@bristol.ac.uk}

\address{Advanced Composites Centre for Innovation and Science, University
of Bristol, BS8 1TR Bristol, UK}

\author{L Boldrin}

\email{l.boldrin@bristol.ac.uk}

\affiliation{Department of Aerospace Engineering, University of Bristol, BS8 1TR
Bristol, UK}

\author{H X Peng}

\email{h.x.peng@bristol.ac.uk}

\address{Advanced Composites Centre for Innovation and Science, University
of Bristol, BS8 1TR Bristol, UK}

\author{C D L Remillat}

\email{c.remillat@bristol.ac.uk}

\affiliation{Department of Aerospace Engineering, University of Bristol, BS8 1TR
Bristol, UK}

\author{S Adhikari}

\email{s.adhikari@swansea.ac.uk}

\affiliation{Multidisciplinary Nanotechnology Centre, University of Swansea, SA2
8PP Swansea, UK}
\begin{abstract}
The temperature-dependent transverse mechanical properties of single-walled
nanotubes are studied using a molecular mechanics approach. The stretching
and bond angle force constants describing the mechanical behaviour
of the $sp^{2}$ bonds are resolved in the temperature range between
0 K and 1600 K, allowing to identify a temperature dependence of the
nanotubes wall thickness. We observe a decrease of the stiffness properties
(axial and shear Young's modulus) with increasing temperatures, and
an augmentation of the transverse Poisson's ratio, with magnitudes
depending on the chirality of the nanotube. Our closed-form predictions
compare well with existing \textcolor{black}{Molecular Dynamics s}imulations. 
\end{abstract}
\maketitle
Carbon nanotubes-based composites, both based on metal/ceramics \cite{ZGD+natmat03},
or polymer matrix \cite{biercuk:2767}, undergo substantial thermal
loading during both the manufacturing phase and their operational
life. This specific aspect of their multifunctional behaviour has
generated a substantial amount of activity in the identification of
the thermal properties of nanotubes, with particular emphasis on the
thermal conductivity \cite{JC+nanotechnology00,JH+appphysA02,CQS+03,KY+prl04,CG+jmps06}.
The dependence of the Young's modulus and tensile strength of single-walled
carbon nanotubes (SWCNTs) has been investigated using various MD techniques
\cite{PhysRevB.67.115407,2006,ZCLSHSapl06,HSY+nanotechnology06} and
hybrid atomistic-Finite Element techniques \cite{LTTXWphysleta07,CX+jrpc09}.
The importance of the interplay between the nanotube and the polymer
matrix during different environmental temperatures has been highlighted
by Wei from\textcolor{black}{{} Monte Carlo/Molecular}\textbf{\textcolor{red}{{}
}}\textcolor{black}{Dynamics (}MC/MD) simulations on a periodic assembly
of a PE-(19,0) CNT nanocomposite \cite{wei:093108}. The use of simulations
to predict the temperature dependence of the nanotubes mechanical
properties is a necessity, because of the difficulties involved in
obtaining direct experimental measurements in different thermal environments.
However, all the predictive tools so far available to simulate and
design the stiffness of CNT-based composites are not based on analytical
formulas that would facilitate the work of the material scientist
and engineer. The existing atomistic-continuum models describing the
elastic properties of carbon nanotubes in closed form \cite{TCaHG03,SLLJphysrevb04}
have their bonds force constant formulated only for room temperature.
In this work we propose a compact formulation that express the relation
between the $sp^{2}$ C-C bond force constants and the surrounding
temperature of the nanotube, providing a functional description of
the coupled thermal and mechanical linear elastic properties of SWCNTs.

In the following description of the model, we will use the suffix
$^{T}$ to indicate the temperature-dependent force constants, while
the suffix $^{o}$ will indicate a force constant term referred to
the room temperature ($T_{0}=300\, K$). Adopting in part the \textcolor{black}{Universal
Force Field (UFF) nomenclatur}e \cite{RAK+92}, the stretching force
constant related to the calculation of the harmonic harmonic potential
energy associated to the stretching deformation is expressed as:

\begin{equation}
k_{IJ}^{T}=664.12\,\frac{Z_{I}^{*}Z_{J}^{*}}{r_{IJ}^{3}}\label{eq:stretching_Badger}\end{equation}

Where $Z_{I}^{*}$ and $Z_{J}^{*}$ are the effective electronic charges
(1.914 electron units). At room temperature, the equilibrium length
$r_{IJ}^{0}$ between the atoms $I$ and $J$ is defined as 0.142
$nm$, with the stretching constant (\ref{eq:stretching_Badger})
assuming the value of $6.52\,10^{-7}N\, nm^{-1}$, which is consistent
with the force constant used in the AMBER model \cite{WDCea95}. The
bending force constant $k_{IJK}^{T}$ related to the harmonic potential
energy associated to the bond bending angle is the second partial
derivative of the bending energy functional \cite{RAK+92}:

\begin{equation}
k_{IJK}^{T}=\left(\frac{\partial^{2}E_{\theta}}{\partial\theta^{2}}\right)=\frac{\overline{\beta}\, Z_{I}^{*}Z_{K}^{*}}{r_{IK}^{5}}\, r_{IJ\,}\, r_{JK}\left\{ 3\, r_{IJ\,}r_{JK}\left[1-\left(\cos\theta\right)^{2}\right]-r_{IK}^{2}\,\cos\theta\right\} \label{eq:kijk_deriv}\end{equation}

Where the distance between the atoms $I$ and $K$ is expressed as
$r_{IK}=r_{IJ}^{2}+r_{JK}^{2}-2\, r_{IJ\,}r_{JK}\cos\theta$. At room
temperature, the bond angle is equal to $\theta=120\, C^{o}$. The
term $\overline{\beta}$ is equal to $332.6/r_{IJ}/r_{jk}$ to obtain
at room temperature a value of $k_{IJK}^{T}$ consistent with the
one provided by the AMBER force, and used by several authors in atomistic-continuum
mechanics models ($8.76\,10^{-10}\, N\, nm/rad^{2}$). The torsional
potential energy between bonds $IJ$ and $KL$ is approximated as
the summation of three cosine Fourier terms based on $\omega$ (the
angle between the $IL$ axis and the $IJK$ plane):

\begin{equation}
E_{\omega}=K_{IJKL}^{T}\sum_{n=0}^{2}C_{n}\cos n\omega_{IJKL}\label{eq:E_omega}\end{equation}

At room temperature the constant $K_{IJKL}^{T}$ is assumed equal
as $2.78\,10^{-10}N/nm/rad^{2}$, consistent with the value used by
several Authors \cite{LCTSWijss03,Odegard20031671}. Due to the very
low sensitivity of the torsional constant versus the temperature,
in our model we consider the term $K_{IJKL}^{T}$ as constant with
the different thermal environments, and will be indicated as $K_{IJKL}^{0}$
in the rest of the work. Following Chen \emph{et al} \cite{CX+jrpc09},
the bond angle $\theta$ is also assumed constant at the room temperature
value ($\theta_{0}=2\,\pi/3$). The thermal dependence of the force
constants (\ref{eq:stretching_Badger}) and (\ref{eq:kijk_deriv})
is provided by the thermal variation of the bond length through an
equivalent coefficient of thermal expansion (CTE) $\alpha$:

\begin{equation}
r_{IJ}=^{0}r_{IJ}\left(1+\alpha\,\Delta T\right)\label{eq:r_T_ij}\end{equation}

Where $\Delta T=T-T_{0}$. In this work, we consider as CTE for the
bond the coefficient of thermal expansion of suspended graphene sheets
calculated using a non-uniform Green's function approach by Jiang
\emph{et al} \cite{JJW+physrevB09}. At room temperature, the coefficient
of thermal expansion is equal to -6 X $10^{-6}\, K^{-1}$, 14 \% lower
than the experimental value measured by Bao \emph{et al} \cite{Bao+09}.
The CTE assumes a value of -3.2 X $10^{-6}\, K^{-1}$ at 20 K, to
decrease to a minimum of -1.25 X $10^{-6}\, K^{-1}$ at \textasciitilde{}
80 K. After this temperature, the coefficient of thermal expansion
increases monotonically, although it remains negative up to 650 K.
At $T=1550\, K$, $\alpha$ brings a value of 5.0 X $10^{-6}\, K^{-1}$.
Assuming $r_{IJ}=r_{JK}$, using (\ref{eq:r_T_ij}) in (\ref{eq:stretching_Badger})
and (\ref{eq:kijk_deriv}), it is possible to express the harmonic
potentials related to the stretching and bending energy respectively
with the variation of the environmental temperature.

\textcolor{black}{The equivalent mechanical behaviour of the C-C bond
is represented at this stage by equating the harmonic potentials with
the correspondent strain energies deformations under axial, bending
and torsion of a structural beam element \cite{LCTSWijss03}}\textbf{\textcolor{red}{:}}

\textcolor{black}{\begin{equation}
\begin{array}{ccc}
\frac{k_{IJ}^{T}}{2}\left(\Delta r\right)^{2}=\frac{EA}{2\, r_{IJ}}\left(\Delta r\right)^{2}, & \frac{K_{IJKL}^{T}}{2}\left(\Delta\varphi\right)^{2}=\frac{GJ}{2\, r_{IJ}}\left(\Delta\varphi\right)^{2}, & \frac{k_{IJK}^{T}}{2}\left(\Delta\theta\right)^{2}=\frac{EI}{2\, r_{IJ}}\frac{4+\Phi}{1+\Phi}\left(\Delta\theta\right)^{2}\end{array}\label{eq:equivalence_ene}\end{equation}
}

\textcolor{black}{In (\ref{eq:equivalence_ene}), $E$ and $G$ are
the equivalent Young's and shear modulus of the material representing
the C-C bond, $I=\pi\, d^{4}/32$ is the polar inertia moment of the
bond beam (considered having a circular cross-section of diameter
$d$, equal to the thickness). Differently from other approaches used
in open literature \cite{KITaPP05}, we adopt in (\ref{eq:equivalence_ene})
a Timoshenko beam having deep shear cross-deformation behaviour as
representative beam model, to consider more realistic distributions
of thickness and equilibrium length existing in graphene and nanotubes
\cite{YH06,FSSA08,FS09,SF+10nano}. The effect due to the shear-induced
cross section deformation is provided by the constant $\Phi=12EI/G/A_{s}/r_{IJ}^{2}$
\cite{Przemienicki_1968,FS09}, where $A_{s}=\pi d^{2}/4/F_{s}$ is
the reduced cross section of the beam by the shear correction term
$F_{s}$ depending on the Poisson's ratio $\nu$ of the equivalent
C-C bond material \cite{TK74}:}

\textcolor{black}{\begin{equation}
F_{s}=\frac{6+12\nu+6\nu^{2}}{7+12\nu+4\nu^{2}}\label{eq:F_s}\end{equation}
}

\textcolor{black}{The equivalence between the harmonic potentials
and the beam strain energies in (\ref{eq:equivalence_ene}) leads
to the following set of equations:}

\textcolor{black}{\begin{equation}
\begin{array}{ccc}
E=\frac{4\, k_{IJ}^{T}r_{IJ}}{\pi\, d^{2}}, & G=\frac{32\, K_{IJKL}^{T}\, r_{IJ}}{\pi\, d^{4}}, & k_{IJK}^{T}=EI\frac{4+\Phi}{r_{IJ}\left(1+\Phi\right)}\end{array}\label{eq:syst_equivalence}\end{equation}
}

\textcolor{black}{Inserting (\ref{eq:F_s}) into the definition of
the shear constant $\Phi$, and solving for} $E$ in (\ref{eq:syst_equivalence}),
we obtain the following nonlinear expression in $d$ and $\nu$ for
a specific temperature $T$:

\begin{equation}
k_{IJK}^{T}=\frac{\pi\, k_{IJ}^{T}\, d^{2}\left(448\,\pi^{2}K_{IJKL}^{T}\, r_{IJ}^{2}+384\,\pi^{2}K_{IJKL}^{T}\, r_{IJ}^{2}\,\nu+9\, k_{IJ}^{T}\pi^{2}d^{4}\nu+9k_{IJ}^{T}\pi^{2}d^{4}\right)}{16\,\pi\left(112\,\pi^{2}r_{IJ}^{2}\, K_{IJKL}^{T}+96\pi^{2}r_{IJ}^{2}\, K_{IJKL}^{T}\nu+9\, k_{IJ}^{T}\pi^{2}d^{4}\nu+9k_{IJ}^{T}\pi^{2}d^{4}\right)}\label{eq:kijk_nonlinear}\end{equation}

We impose the \textcolor{black}{additional conditi}on that the equivalent
material of the C-C bond beam behaves as \emph{isotropic}, i.e., with
no directional preference in its mechanical response. The isotropic
condition $G=E/2/\left(1+\nu\right)$, together with relation (\ref{eq:kijk_nonlinear})
constitutes a system of nonlinear equations which can be solved with
traditional methods, such as\textcolor{black}{{} the Ma}rquardt algorithm.
The solution of the system yields a unique value of the thickness
$d$ for a given temperature $T$. At room temperature, the thickness
identified using this approach yields a value of 0.084 nm, equal to
the one found for single wall and nanotube bundles in \cite{FSSA08}.
We observe a low sensitivity of the thickness corresponding to the
isotropic condition versus the temperature, with a minimum value of
0.0835 nm at 1600 $K$, and a maximum value of 0.0841 nm for $T=470\, K$.
The temperature dependence of the thickness can be described using
a polynomial curve of the $9^{th}$ order, as shown in Table \ref{Flo:tab_fittings}.
It is interesting to notice that the C-C bond behaves mechanically
as if made of a quasi-zero Poisson's ratio behaviour, like in natural-occurring
cork \cite{LJGMA97b}. The 0.084 nm value at room temperature is also
consistent with the 0.084 nm found by Kudin \emph{et al} \cite{Kudin2001prb},
and 0.074 nm identified by Tu and Ou-Yang \cite{ZTaZO02}. From simulations
using the MM3 potential, Batra and Sears have found an equivalent
thickness of 0.043 nm for different chiral configurations \cite{BRCSAmsmse07},
although a value of 0.1 nm was observed in nanotubes undergoing bending
and breathing vibration modes \cite{BRCGSSasmejap08}. Zhang and Shen
identify thickness of 0.088 nm and 0.087 nm for (17,0) and (21,0)
nanotubes respectively, while for (10,10) and (12,12) the thickness
calculated is 0.067 nm \textcolor{black}{\cite{ZCLSHSapl06}. The
thickness values calculated with our method do compare well with the
0.080 nm of Chen and Cao \cite{ChenCaonanotechnology06}, although
they are higher than the 0.066 nm of Yakobson }\textcolor{black}{\emph{et
al}}\textcolor{black}{{} \cite{PhysRevLett.76.2511}, and 0.60 nm of
Vodenitcharova and Zhang \cite{PhysRevB.68.165401}. }

\begin{figure}[p]
\includegraphics[scale=0.6]{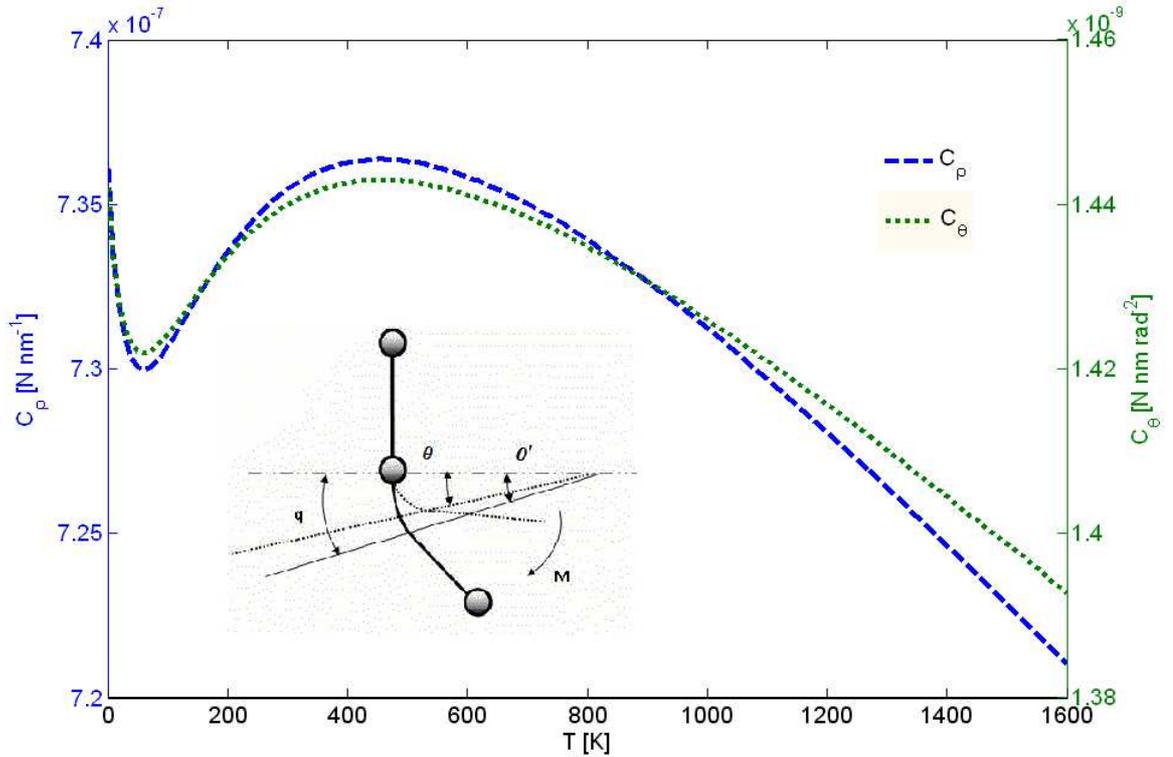}

\caption{The variation of the stretching constant $C_{\rho}$ and hinging constant
$C_{\theta}$ over the temperature. The hinging constant is calculated
considering the change in angular deformation $\theta-\theta^{,}=\Delta\theta=\int_{0}^{q}\frac{M}{EI}\, dq=Mq/EI.$
Under hinging, only the initial portion $q\simeq r_{IJ}/5$ of the
bond length deforms by rotation, while the rest of the bond deforms
as a rigid body \cite{MIGKEEcost96}. The hinging constant is then
defined as $C_{\theta}=M/\Delta\theta=EI/q$.}

\label{Flo:cRHO_cTHETA_t} 
\end{figure}

\textcolor{black}{Classical mechanistic theories for the elastic properties
of single-wall carbon nanot}ubes consider the strain energy associated
to the deformation of $sp^{2}$ bonds in terms of stretching and bond
angle constants ($C_{\rho}$ and $C_{\theta}$ respectively \cite{TCaHG03,SLLJphysrevb04}).
For the bond angle constant (moment), we consider that for linear
elasticity the lattice is dominated by hinging deformation \cite{Gillis1984c}.
In cellular structures under general flexural/axial deformations,
the hinging constant can be calculated as $C_{\theta}=EI/q$, where
$q$ is the portion of the C-C bond length undergoing hinging, while
the segment $r_{IJ}-q$ deforms as a rigid body (\cite{MIGKEEcost96},
see also Figure \ref{Flo:cRHO_cTHETA_t}). Using the relations (\ref{eq:syst_equivalence}),
the bending angle constant can be rewritten as $C_{\theta}=k_{IJ}^{T}\, d^{2}r_{IJ}/16/q$.
Considering a ratio $r_{IJ}/q=5$, inserting the numerical values
of the thickness and equilibrium length at room temperature we obtain
a value of 1.44 $nN\, nm\, rad^{-2}$, well in line with the 1.42
$nN\, nm\, rad^{-2}$ used in \cite{TCaHG03,SLLJphysrevb04}. The
stretching constant is calculated as $C_{\rho}=k_{\rho}\, k_{IJ}^{T}$,
where $k_{\rho}=1.128$ to converge to the 0.36 $TPa\, nm^{-1}$ of
the in-plane graphitic surface modulus (chiral index $n\rightarrow\infty$)
\cite{Gillis1984c,JC07}. At room temperature, the stretching constant
identified in this model is equal to 735 $nN\, nm^{-1}$, which compares
well with the 742 $nN\, nm^{-1}$ used by Shen and Li \cite{SLLJphysrevb04}.
\textcolor{black}{Cadelano }\textcolor{black}{\emph{et al \cite{PhysRevLett.102.235502}}}\textcolor{black}{{}
have identified a value of the surface modulus of graphene-type systems
equal to 0.312 $TPa\, nm^{-1}$ at zero temperature damped dynamics
using tight binding calculations and continuum elasticity. Using our
approach, we observe a value of 0.332 $TPa\, nm^{-1}$ at zero temperature,
6 \% higher. For both the bending and stretching constants, we observe
a very low dependence versus the environmental temperature between
$370\, K<T<570\, K$. The s}tretching constant then decreases with
increasing temperatures with an approximate rate of 0.015 $nN\, nm^{-1}K^{-1}$,
a behaviour similar to the one of the $C_{\theta}$ constant, having
the latter a decrease rate of $4.9\,10^{-5}\, nN\, nm\, rad^{-2}K^{-1}$
between 600 K and 1600 K. At low temperatures, both the force and
bending constants decrease from values equal to 99.2 \% of the maximum
$C_{\theta}$ and $C_{\rho}$ at 1 K to a minimum at 70 K corresponding
to the 98.6 \% of the maximum of the constants at 450 K. Similarly
to the C-C bond thickness, the force stretching and bending constants
can be expressed in polynomial terms, as illustrated in Table \ref{Flo:tab_fittings}.

\begin{figure}[p]
\includegraphics{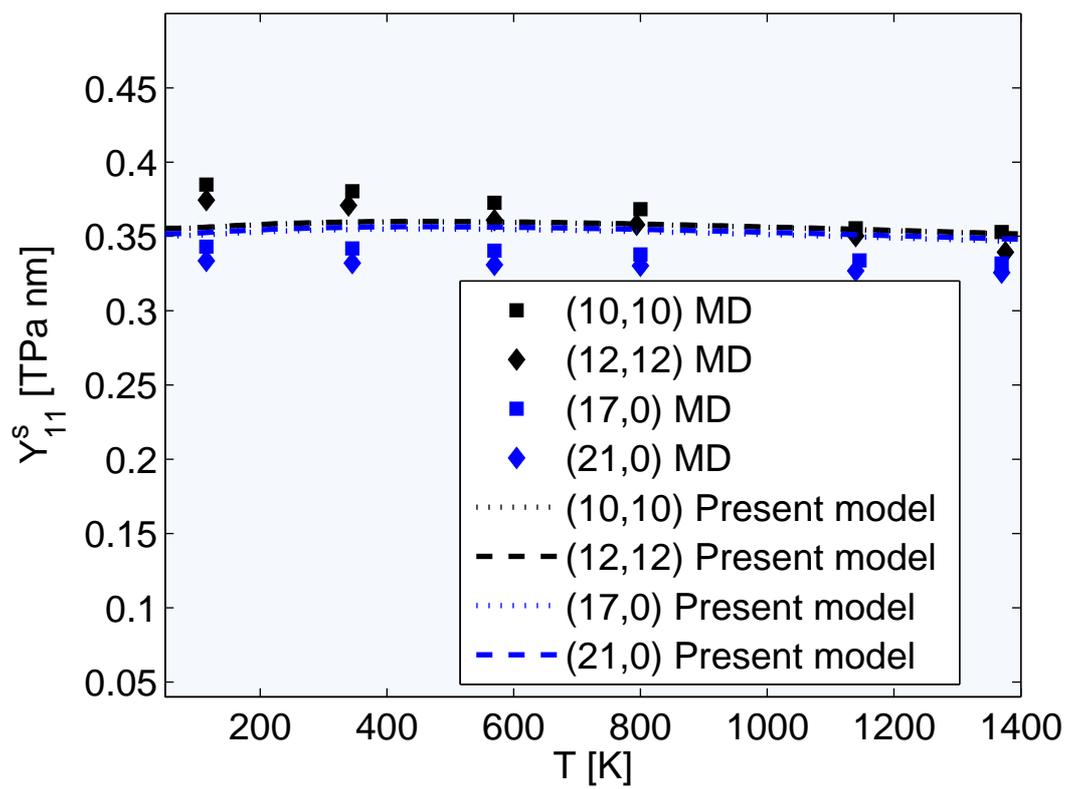}

\caption{Comparison between the surface Young's modulus of different zigzag
and armchair SWCNTs and the MD simulations from \cite{ZCLSHSapl06}}

\label{Flo:alpl_89_comp} 
\end{figure}

\begin{figure}[p]
\includegraphics{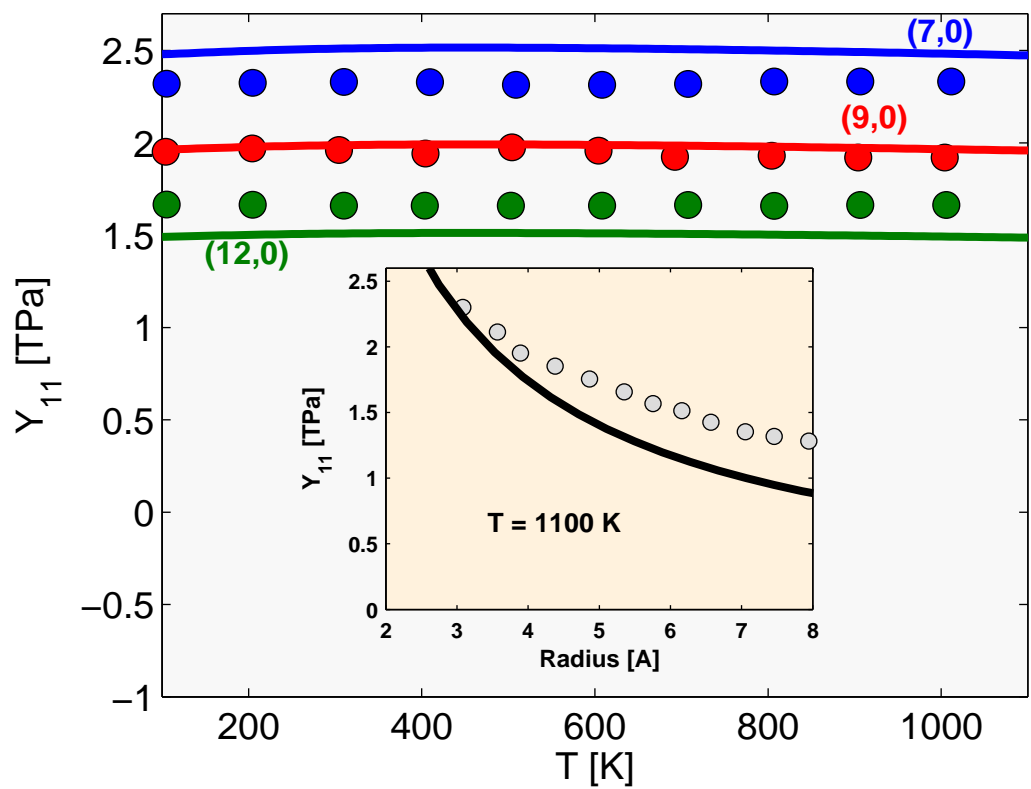}

\caption{The effective Young's modulus modulus for $\left(n,0\right)$ SWCNTs.
The results from the proposed method (continuous line) are compared
with the ones from Hsieh and co-Authors \cite{HSY+nanotechnology06}
(circular dots).}

\label{Flo:comp_Y11} 
\end{figure}

\begin{figure}[p]
\includegraphics{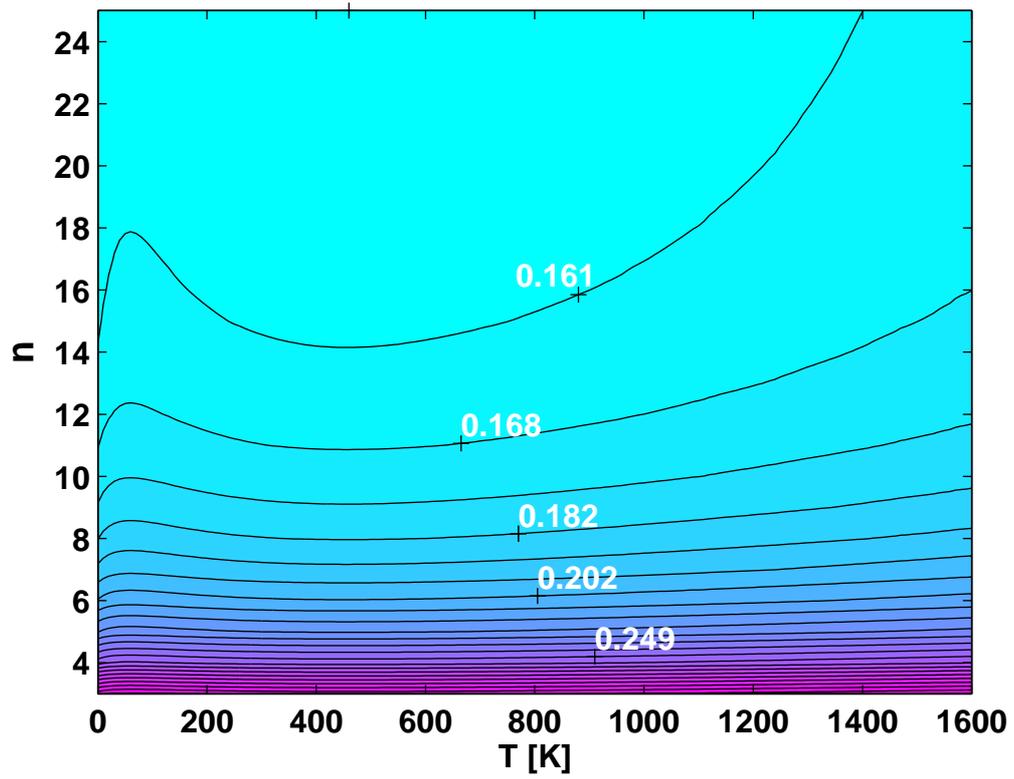}

\caption{Map of the transverse Poisson's ratio $\nu_{21}$ for $\left(n,0\right)$
tubes versus the the chiral index $n$ and temperature.}

\label{Flo:Fig_nu21_T_n_zig}
\end{figure}

The temperature-dependent constants $C_{\rho}$ and $C_{\theta}$
can now be used in atomistic-continuum approaches illustrated in open
literature to model in a closed-form solution the mechanical properties
of SWCNTs under different temperature conditions. In this work we
have used the formulation proposed by Shen and Li \cite{SLLJphysrevb04}
to simulate the axial and shear stiffness, as well as the Poisson's
ratio of the nanotubes with armchair and zigzag geometry. We compare
the surface Young's modulus $Y_{11}^{s}$ calculated in analytical
form against the results obtained through MD simulations using a REBO
potential by Zhang and Shen \cite{ZCLSHSapl06} (Figure \ref{Flo:alpl_89_comp}).
\textcolor{black}{Our predictions are well in line with the 0.352
$TPa\, nm^{-1}$ of Tu and Ou-Yang \cite{ZTaZO02} and Pantano }\textcolor{black}{\emph{et
al}}\textcolor{black}{{} \cite{PhysRevLett.91.145504}, although 3.5
\% higher than Li and Chou \cite{LCTSWijss03} and Wang and co-Authors
\cite{PhysRevLett.95.105501}. In terms of longitudinal modulus $Y_{11}^{l}$
(defined as $Y_{11}^{s}/d$), our} predictions tend to overestimate
the MD simulations in \cite{ZCLSHSapl06}. For example, our axial
moduli for (17,0) nanotubes are 9 \% higher than the molecular dynamics
simulations at $T=300\, K$, while a at a temperature of 1000 K our
overestimate is around 11 \%. The effective Young's modulus $Y_{11}=Y_{11}^{s}/\left(R/2\right)$(where
$R$ is the diameter of the nanotube) has been also derived by Hsieh\emph{
et al} \cite{HSY+nanotechnology06} from the vibration amplitude of
$(n,0)$ clamped-free nanotubes simulated with a Tersoff-Brenner potential
at temperatures ranging between 0 K and 2000 K. The MD simulations
in \cite{HSY+nanotechnology06} are least-squares fitted to an asymptotic
value of the graphene Young's modulus equal to 1.2 TPa found by \cite{HE+prl98,GVL00}
at a not specified temperature, 20 \% higher than the experimental
value found by Lee \emph{et al} for a thickness of 0.34 nm \cite{Lee2008sc}.
A general comparison of the results from \cite{HSY+nanotechnology06}
against our simulations is presented in Figure \ref{Flo:comp_Y11},
showing a general agreement between trends related to the nanotube
radius at $T=1100\, K$, and within the temperature range 50 K - 1100
K. We observe a good convergence between MD and our simulations especially
for CNT radius lower than 0.5 nm, and excellent agreement (percentage
error less than 4 \%) for (7,0) SWCNTs over the temperature range
considered. Hsieh and co-Authors record a steep decrease in terms
of axial stiffness above 1100 K, corresponding to an abrupt change
of the standard deviation associated to the dynamic tip displacements
of the tube \cite{HSY+nanotechnology06}. We observe also an overestimate
for the shear modulus $G_{12}$ \cite{ZCLSHSapl06}, with discrepancies
around 29 \% and 28 \% at 1000 K and 1200 K respectively. However,
our predictions show a decrease of the longitudinal shear modulus
versus the temperature, opposite from the results derived in \cite{ZCLSHSapl06}.
The transverse Poisson's ratio $\nu_{21}$ increases with the temperature
(Figure \ref{Flo:Fig_nu21_T_n_zig}), although the augmentation is
limited on average to a 3 \% increase between $T=300\, K$ and $T=1200\, K$
for the different chirality and tube diameters. The low sensitivity
of the Poisson's ratio over the temperature is due to the weak dependency
versus the temperature itself of the term $C_{\rho\,}r_{IJ}^{2}/C_{\theta}$,
which is contained both in the numerator and denominator of the Poisson's
ratio expression \cite{SLLJphysrevb04}. Chen \emph{et al.} \cite{CX+jrpc09}
observe also an increase of $\nu_{21}$ with temperature (31 \% of
increase for zizgag tubes of 1.2 nm diameter between room temperature
and $T=1200\, K$). The discrepancy between theirs and our predictions
is due mainly to the coefficient of thermal expansion used for the
C-C bond, being positive in their case for $T>298\, K$, while our
model has a positive CTE for temperatures higher than 650 K \cite{JJW+physrevB09}.
We observe that Zhang and Shen predict a transverse Poisson's ratio
for a (17,0) nanotube at $T=300\, K$ equal to 0.17, in good agreement
with our value of 0.16 \cite{ZCLSHSapl06}.\textcolor{black}{{} We note
that for graphene-type systems, our model predicts at zero temperature
a Poisson's ratio of 0.12, significantly lower than the 0.31 identified
in \cite{PhysRevLett.102.235502}. However, our range of Poisson's
ratios at room temperature for zigzag tubes (Figure \ref{Flo:Fig_nu21_T_n_zig})
is in line with the 0.12 - 0.19 values predicted by Sanchez-Portal
}\textcolor{black}{\emph{et al }}\textcolor{black}{\cite{PhysRevB.59.12678},
the 0.15 of Kudin and co-Authors \cite{Kudin2001prb}, 0.21 of Sears
and Batra \cite{PhysRevB.69.235406}, and 0.24 from Zhou }\textcolor{black}{\emph{et
al}}\textcolor{black}{{} \cite{PhysRevB.62.13692}.}

In summary, our model allows to predict in a compact form the overall
elastic transverse properties of single wall nanotubes over a large
temperature range. The bending and stretching constants identified
with our approach capture the intrinsic hinging/stretching behaviour
of $sp^{2}$ C-C bonds in nanotubes, and relate the stiffness and
Poisson's ratio to the ambient temperature of the nanotube, with 2
\% average decrease of the Young's modulus between the ambient temperature
and 1200 K for a given tube chirality.

\bibliographystyle{apsrev} \bibliographystyle{apsrev}
\bibliography{thermo_CNT_fs_lb_hxp_cdlr_sa_7}

\begin{table}[p]
\begin{tabular}{|c|c|c|c|c|c|c|c|c|c|c|}
\hline 
Quantity  & $p_{1}$  & $p_{2}$  & $p_{3}$  & $p_{4}$  & $p_{5}$  & $p_{6}$  & $p_{7}$  & $p_{8}$  & $p_{9}$  & $p_{10}$\tabularnewline
\hline
\hline 
$d$ X $10^{-5}$ {[}nm{]}  & $-2.256$  & $2.571$  & $12.7$  & $-13.6$  & $-17.78$  & $13.14$  & $14.81$  & $14.13$  & $-23.66$  & $8401$\tabularnewline
\hline 
$C_{\theta}$ X $10^{12}$ $\left[N\, nm\, rad^{-2}\right]$  & $-1.84$  & $2.181$  & $9.427$  & $-9.675$  & $-14.92$  & $12.9$  & $12.47$  & $-11.97$  & $-19.8$  & $1435$\tabularnewline
\hline 
$C_{\rho}$ X $10^{10}$ $\left[N\, nm^{-1}\right]$  & $-5.731$  & $6.702$  & $2873$  & $-29.83$  & $-45.55$  & $35.13$  & $37.83$  & $-36.98$  & $-59.98$  & $7340$\tabularnewline
\hline
\end{tabular}

\caption{Polynomial interpolations for the thickness, stretching and hinging
force constants for the $sp^{2}$ C-C bond lattice versus the temperature.
The polynomials are of the type $y=\sum_{n=1}^{10}p_{n}\widehat{T}^{10-n}$,
where $\widehat{T}=\left(T-\overline{T}\right)/\sigma\left(T\right)$,
with $\overline{T}=800\, K$ and standard deviation $\sigma\left(T\right)=466\, K$.
All fittings have an average $R^{2}$ of 0.991, with 95 \% of confidence
interval.}

\label{Flo:tab_fittings} 
\end{table}

\end{document}